# Thermal Transport in Graphene, Few-Layer Graphene and Graphene Nanoribbons

**Denis L. Nika[1] and Alexander A. Balandin[2,*]**

[1]Department of Physics and Engineering, Moldova State University, Chisinau, MD-2009, Republic of Moldova

[2]Department of Electrical and Computer Engineering and Materials Science and Engineering Program, University of California, Riverside, CA 92521 U.S.A.

**Abstract**  The discovery of unusual heat conduction properties of graphene has led to a surge of theoretical and experimental studies of phonon transport in two-dimensional material systems. The rapidly developing graphene thermal field spans from theoretical physics to practical engineering applications. In this invited review we outline different theoretical approaches developed for describing phonon transport in graphene and provide comparison with available experimental thermal conductivity data. A special attention is given to analysis of the recent theoretical results for the phonon thermal conductivity of graphene and few-layer graphene, the effects of the strain, defects, isotopes and edge scattering on the acoustic phonon transport in these material systems.

## 1. Introduction

Thermal management has become a crucial issue for continuing progress in electronic industry owing to increased levels of dissipated power density and speed of electronic circuits [1]. Self-heating is a major problem in electronics, optoelectronics and photonics [2]. These facts stimulated practical interest in thermal properties of materials. Acoustic phonons are the main heat carriers in a variety of material systems. The phonon and thermal properties of nanostruc-tures are substantially different from those of bulk crystals [3-15]. Semiconductor nanostructures do not conduct heat as well as bulk crystals due to increased phonon - boundary scattering [4-5] as well as changes in the phonon dispersion and density of states (DOS) [3-8]. From the other side, theoretical studies suggested that phonon transport in strictly two-dimensional (2D) and one-dimensional (1D) systems can reveal exotic behavior, leading to infinitely large *intrinsic* thermal conductivity [9-12]. These theoretical results have led to discussions of the validity of Fourier's law in low-dimensional systems [16-17] and further stimulated interest in the acoustic phonon transport in 2D systems.

[*]Corresponding author email address (A.A.B.): balandin@ee.ucr.edu

In this Chapter, we focus on the specifics of the acoustic phonon transport in graphene. The Chapter is mostly based on our original and review papers dedicated to various aspects of heat conduction in graphene [18-28]. After a summary of the basics of thermal physics in nanostructures and experimental data for graphene's thermal conductivity, we discuss, in more detail, various theoretical approaches to calculation of the phonon thermal conductivity in graphene. Special attention is given to the analysis of the most recent theoretical results on the relative contributions of different phonon polarization branches to the thermal conductivity of graphene. The readers interested in the experimental thermal conductivity values of graphene and related materials are referred to a complementary review [18].

## 1. Basics of phonon transport and thermal conductivity

The main experimental technique for investigation of the acoustic phonon transport in a given material system is the measurement of its lattice thermal conductivity [29-30]. The thermal conductivity is introduced through Fourier's law [31-32]:

$$\vec{\phi} = -K\nabla T , \qquad (1)$$

where $\vec{\phi}$ is the heat flux, $\nabla T$ is the temperature gradient and $K = (K_{\alpha\beta})$ is the thermal conductivity tensor. In the isotropic medium, thermal conductivity does not depend on the direction of the heat flow and $K$ is treated as a constant. The latter is valid for the small temperature variations only. In a wide temperature range, thermal conductivity is a function of temperature, i.e. $K \equiv K(T)$. In general, in solid materials heat is carried by phonons and electrons so that $K = K_p + K_e$, where $K_p$ and $K_e$ are the phonon and electron contributions, respectively. In metals or degenerately-doped semiconductors, $K_e$ is dominant due to the large density of free carriers. The value of $K_e$ can be determined from the measurement of the electrical conductivity $\sigma$ via the Wiedemann-Franz law [33]:

$$\frac{K_e}{\sigma T} = \frac{\pi^2 k_B^2}{3e^2},\qquad(2)$$

where $k_B$ is the Boltzmann's constant and $e$ is the charge of an electron. Phonons are usually the main heat carriers in carbon materials. Even in graphite, which has metal-like properties [34], the heat conduction is dominated by acoustic phonons [35]. This fact is explained by the strong covalent $sp^2$ bonding, resulting in high in-plane phonon group velocities and low crystal lattice unharmonicity for in-plane vibrations.

The phonon thermal conductivity can be written as

$$K_p = \Sigma_j \int C_j(\omega) \upsilon_{x,j}(\omega) \upsilon_{x,j}(\omega) \tau_j(\omega) d\omega,\qquad(3)$$

where summation is performed over the phonon polarization branches $j$, which include two transverse acoustic branches and one longitudinal acoustic branch, $\upsilon_{x,j}$ is the projection of the phonon group velocity $\vec{\upsilon}_j = d\omega_j / d\vec{q}$ on the $X$-axis for the $j$th branch, which, in many solids, can be approximated by the sound velocity, $\tau_j$ is the phonon relaxation time, $C_j = \hbar \omega_j \partial N_0(\hbar \omega_j / k_B T) / \partial T$ is the contribution to heat capacity from the $j$th branch, and $N_0(\frac{\hbar \omega_j}{k_B T}) = [\exp(\frac{\hbar \omega_j}{k_B T}) - 1]^{-1}$ is the Bose-Einstein phonon equilibrium distribution function. The phonon mean-free path (MFP) $\Lambda$ is related to the relaxation time through the expression $\Lambda = \tau \upsilon$. In the relaxation-time approximation (RTA), various scattering mechanisms, which limit the MFP, usually considered as additive, i.e. $\tau_j^{-1} = \sum_i \tau_{i,j}^{-1}$, where $i$ denotes scattering mechanisms. In typical solids, acoustic phonons, which carry the bulk of heat, are scattered by other phonons, lattice defects, impurities, conduction electrons, and interfaces [36-39].

In ideal crystals, i.e. crystals without defects or rough boundaries, $\Lambda$ is limited by the phonon - phonon scattering due to the crystal lattice anharmonicity. In this case, thermal conductivity is referred to as intrinsic. The anharmonic phonon interactions, which lead to the finite thermal conductivity in three dimensions, can be described by the Umklapp processes [36]. The Umklapp scattering rates depend on the Gruneisen parameter $\gamma$, which determines the degree of the lattice anharmonicity [36-37]. Thermal conductivity is extrinsic when it is mostly limited by the extrinsic effects such phonon – boundary or phonon – defect scattering.

In nanostructures, the phonon energy spectra are quantized due to the spatial confinement of the acoustic phonons. The quantization of the phonon energy spectra, typically, leads to decreasing phonon group velocity. The modification of the phonon energies, group velocities and density of states, together with phonon scattering from boundaries affect the thermal conductivity of nanostructures. In most of cases, the spatial confinement of acoustic phonons results in a reduction of the phonon thermal conductivity [40-41]. However, in some cases, the thermal conductivity of nanostructures embedded within the acoustically hard barrier layers can be increased via spatial confinement of acoustic phonons [6-7, 10, 42].

The phonon boundary scattering can be evaluated as [39]

$$\frac{1}{\tau_{B,j}} = \frac{\upsilon_{x,j}}{D} \frac{1-p}{1+p},\qquad(4)$$

where $D$ is the nanostructure or grain size and $p$ is the specularity parameter defined as a probability of specular scattering at the boundary. The momentum-conserving specular scattering ($p$=1) does not add to thermal resistance. Only diffuse phonon scattering from rough interfaces ($p$→0), which changes the phonon momen-tum, limits the phonon MFP. The commonly used expression for the phonon specularity is given by [39, 43-44]

$$p(\lambda) = \exp(-\frac{16\pi^2\eta^2}{\lambda^2}),$$  (5)

where $\eta$ is the root mean square deviation of the height of the surface from the reference plane and $\lambda$ is the phonon wavelength..

When the phonon - boundary scattering is dominant, the thermal conductivity scales with the nanostructure or grain size $D$ as $K_p \sim C_p\upsilon\Lambda \sim C_p\upsilon^2\tau_B \sim C_p\upsilon D$. In nanostructures with $D << \Lambda$, the thermal conductivity dependence on the physical size of the structure becomes more complicated due to the strong quantization of the phonon energy spectra [6, 40, 42]. The specific heat $C_p$ depends on the phonon density of states, which leads to different $C_p(T)$ dependences in three-dimensional (3D), two-dimensional and one-dimensional systems, and reflected in $K(T)$ dependence at low $T$ [36, 39]. In bulk at low $T$, $K(T) \sim T^3$ while it is $K(T) \sim T^2$ in 2D systems.

The thermal conductivity $K$ defines how well a given material conducts heat. The thermal diffusivity, $\alpha$, defines how fast the material conducts heat. It is given by the expression

$$\alpha = \frac{K}{C_p\rho_m},$$  (6)

where $\rho_m$ is the mass density. Many experimental techniques directly measure thermal diffusivity rather than thermal conductivity.

## 2. Experimental data for thermal conductivity of graphene and few-layer graphene

The first measurements of heat conduction in graphene [19-22, 45-46] were carried out at the University of California – Riverside (see figure 1). The experimental study was made possible by the development of the optothermal technique. The experiments were per-formed with the large-area suspended graphene layers exfoliated from the high-quality Kish and highly ordered pyrolytic graphite. It was found that the thermal conductivity varies in a wide range and can exceed that of the bulk graphite, which is ~2000 W/mK at room temperature (RT). It was also determined that the electronic contribution to heat conduction in the un-gated graphene near RT is much smaller than that of phonons, i.e. $K_e << K_p$. The phonon MFP in graphene was estimated to be on the order of 800 nm near RT [20].

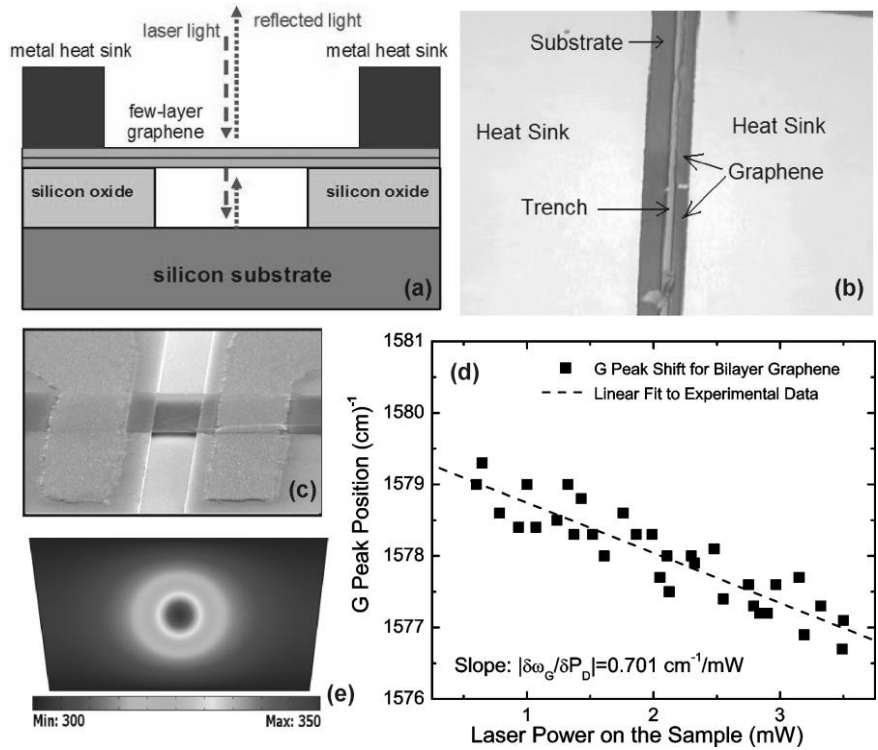

**Fig. 1.** Illustration of optothermal micro-Raman measurement technique developed for investigation of phonon transport in graphene. (a) Schematic of the thermal conductivity measurement showing suspended FLG flakes and excitation laser light. (b) Optical microscopy images of FLG attached to metal heat sinks. (c) Colored scanning electron microscopy image of the suspended graphene flake to clarify typical structure geometry. (d) Experimental data for Raman G-peak position as a function of laser power, which determines the local temperature rise in response to the dissipated power. (e) Finite-element simulation of temperature distribution in the flake with the given geometry used to extract the thermal conductivity. Figure is after Ref. [22] reproduced with permission from the Nature Publishing Group.

Several independent studies, which followed, also utilized the Raman optothermal technique but modified it via addition of a power meter under the suspended portion of graphene. It was found that the thermal conductivity of suspended high-quality chemical vapour deposited (CVD) graphene exceeded ~2500 W/mK at 350 K, and it was as high as $K \approx 1400$ W/mK at 500 K [47]. The reported value was also larger than the thermal conductivity of bulk graphite at RT. Another Raman optothermal study with the suspended graphene found the thermal conductivity in the range from ~1500 to ~5000 W/mK [48]. Another group that repeated the Raman-based measurements found $K \approx 630$ W/mK for a suspended graphene membrane [49]. The differences in the actual temperature of graphene under laser heating, strain distribution in the suspended graphene of various sizes and geometries can explain the data variation.

Another experimental study reported the thermal conductivity of graphene to be ~1800 W/mK at 325 K and ~710 W/mK at 500 K [50]. These values are lower than that of bulk graphite. However, instead of measuring the light absorption in graphene under conditions of their experiment, the authors of Ref. [50] assumed that the optical absorption coefficient should be 2.3%. It is known that due to many-body effects, the absorption in graphene is the function of wavelength $\lambda$, when $\lambda > 1$ eV [51-53]. The absorption of 2.3% is observed only in the near-infrared at ~1 eV. The absorption steadily increases with decreasing $\lambda$ (increasing energy). The 514.5-nm and 488-nm Raman laser lines correspond to 2.41 eV and 2.54 eV, respectively. At 2.41 eV the absorption is about $1.5 \times 2.3\% \approx 3.45\%$ [51]. The value of 3.45% is in agreement with the one reported in another independent study [54]. Replacing the assumed 2.3% with 3.45% in the study reported in Ref. [50] gives ~2700 W/mK at 325 K and 1065 W/mK near 500 K. These values are higher than those for the bulk graphite and consistent with the data reported by other groups [47, 54], where the measurements were conducted by the same Raman optothermal technique but with the measured light absorption.

The data for suspended or partially suspended graphene is closer to the intrinsic thermal conductivity because suspension reduces thermal coupling to the substrate and scattering on the substrate defects and impurities. The thermal conductivity of fully supported graphene is smaller. The measurements for exfoliated graphene on $SiO_2$/Si revealed in-plane $K \approx 600$ W/mK near RT [55]. Solving the Boltzmann transport equation (BTE) and comparing with their experiments, the authors determined that the thermal conductivity of free graphene should be ~3000 W/mK near RT.

Despite the noted data scatter in the reported experimental values of the thermal conductivity of graphene, one can conclude that it is very large compared to that for bulk silicon ($K$=145 W/mK at RT) or bulk copper ($K$=400 W/mK at RT) – important materials for electronic applications. The differences in $K$ of graphene can be attributed to variations in the graphene sample lateral sizes (length and width), thickness non-uniformity due to the mixing between single-layer and few-layer graphene, material quality (e.g. defect concentration and surface contaminations), grain size and orientation, as well as strain distributions. Often the reported thermal conductivity values of graphene corresponded to different sample temperatures $T$, despite the fact that the measurements were conducted at ambient temperature. The strong heating of the samples was required due to the limited spectral resolution of the Raman spectrometers used for temperature measurements. Naturally, the thermal conductivity values determined at ambient but for the samples heated to $T$~350 K and $T$~600 K over a substantial portion of their area would be different and cannot be directly compared. One should also note that the data scatter for thermal conductivity of carbon nanotubes (CNTs) is much larger than that for graphene. For a more detail analysis of the experimental uncertainties the readers are referred to a comprehensive review [18].

The phonon thermal conductivity undergoes an interesting evolution when the system dimensionality changes from 2D to 3D. This evolution can be studied with the help of suspended few-layer graphene (FLG) with increasing thickness $H$ – number of atomic planes $n$. It was reported in Ref. [22] that thermal conductivity of suspended uncapped FLG decreases with increasing $n$ approaching the bulk graphite limit (see figure 2). This trend was explained by considering the intrinsic quasi-2D crystal properties described by the phonon Umklapp scattering [22]. As $n$ in FLG increases – the phonon dispersion changes and more phase-space states become available for phonon scattering leading to thermal conductivity decrease. The phonon scattering from the top and bottom boundaries in suspended

FLG is limited if constant $n$ is maintained over the layer length. The small thickness of FLG ($n$<4) also means that phonons do not have transverse cross-plane component in their group velocity leading to even weaker boundary scattering term for the phonons. In thicker FLG films the boundary scattering can increase due to the non-zero cross-plane phonon velocity component. It is also harder to maintain the constant thickness through the whole area of FLG flake. These factors can lead to a thermal conductivity below the graphite limit. The graphite value is recovered for thicker films.

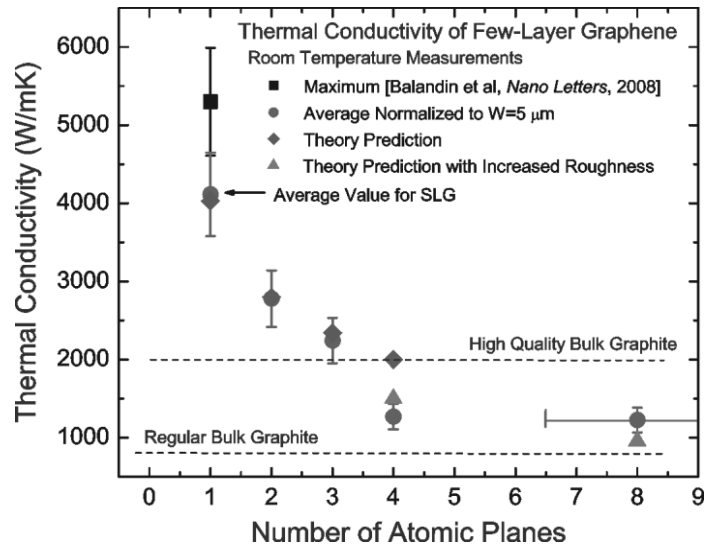

**Fig. 2.** Measured thermal conductivity as a function of the number of atomic planes in FLG. The dashed straight lines indicate the range of bulk graphite thermal conductivities. The blue diamonds were obtained from the first-principles theory of thermal conduction in FLG based on the actual phonon dispersion and accounting for all allowed three-phonon Umklapp scattering channels. The green triangles are Callaway–Klemens model calculations, which include extrinsic effects characteristic for thicker films. Figure is after Ref. [22] reproduced with permission from the Nature Publishing Group.

The experimentally observed evolution of the thermal conductivity in FLG with $n$ varying from 1 to $n$~4 [22] is in agreement with the theory for the crystal lattices described by the Fermi-Pasta-Ulam Hamiltonians [56]. The molecular-dynamics (MD) calculations for graphene nanoribbons with the number of planes $n$ from 1 to 8 [57] also gave the thickness dependence of the thermal conductivity in agreement with the UC Riverside experiments [22]. The strong re-duction of the thermal conductivity as *n* changes from 1 to 2 is in line with the earlier theoretical predictions [58]. In another reported study, the Boltzmann transport equation was solved under the assumptions that in-plane interactions are described by Tersoff potential while the Lennard-Jones potential models interactions between atoms belonging to different layers [59-60]. The obtained results suggested a strong thermal conductivity decrease as *n* changed from 1 to 2 and slower decrease for *n*>2.

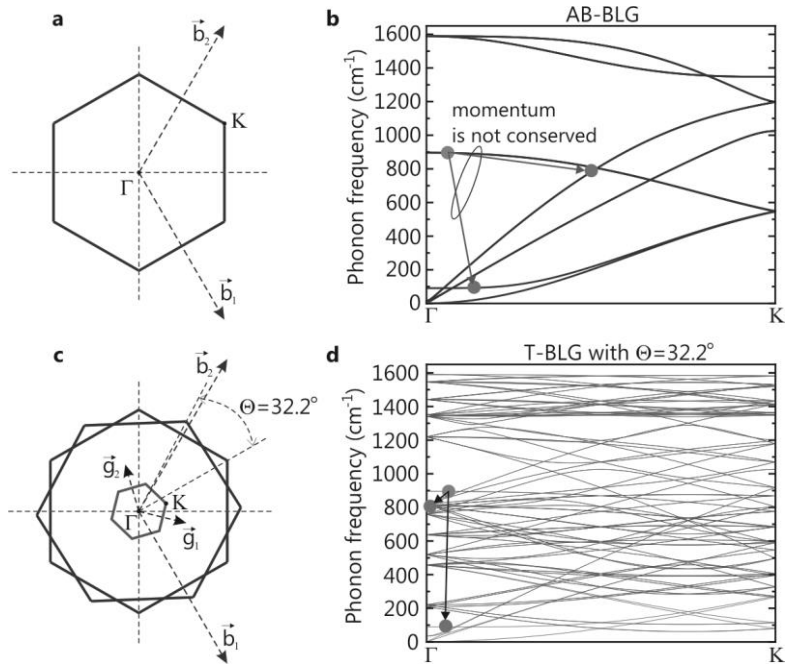

**Fig. 3.** (a-b) Brillouin zone and calculated phonon dispersions for Bernal-stacked bilayer graphene and (c-d) Brillouin zone and calculated phonon dispersion for twisted bilayer graphene. The twist angle in the calculation was assumed to be 32.2º. Note that new channels of phonon relaxation appear in twisted graphene: the normal decay of the phonon with $\omega = 900\,\mathrm{cm}^{-1}$ (blue point) into two phonons with $\omega' = 800\,\mathrm{cm}^{-1}$ and $\omega'' = 100\,\mathrm{cm}^{-1}$ (red points) is allowed by the momentum conversation law in TBLG and is not allowed in AB-BLG (panel (d)). Figure is after Ref. [62] reproduced with permission from the Royal Society of Chemistry.

The thermal conductivity dependence on the FLG is entirely different for the encased FLG where thermal transport is limited by the acoustic phonon scattering from the top and bottom boundaries and disorder. The latter is common when FLG is embedded between two layers of dielectrics.

**Table 1.** Thermal conductivity of graphene and graphene nanoribbons: experimental data.

| Sample | K (W/mK) | Method | Description | Ref |
|---|---|---|---|---|
| SLG | ~2000 – 5000 | Raman opto-thermal | suspended; exfoliated | 19,20 |
| | ~2500 | Raman opto-thermal | suspended; chemical vapor deposition (CVD) grown | 47 |
| | ~1500-5000 | Raman opto-thermal | suspended; CVD grown | 48 |
| | 600 | Raman opto-thermal | suspended; exfoliated; T ~ 660 K | 49 |
| | 600 | electrical | supported; exfoliated; | 55 |
| | 310 - 530 | electrical self-heating | exfoliated and chemical vapor deposition grown; $T$~1000 K. | 65 |
| | 2778 ± 569 | Raman opto-thermal | suspended, CVD-grown | 62 |
| | ~ 1700 | electrical self heating | suspended; CVD-grown; flake length ~ 9 μm; strong length dependence | 66 |
| Bilayer graphene | ~1900 | Raman opto-thermal | Suspended; T~320 K | 62 |
| | 560-620 | electrical self-heating | suspended; polymeric residues on the surface. | 67 |
| Twisted bi-layer | ~1400 | Raman opto-thermal | Suspended; T~320 K | 62 |
| FLG | 1300 - 2800 | Raman opto-thermal | suspended; exfoliated; n=2-4 | 22 |
| | 50 - 970 | heat-spreader method | FLG, encased within $SiO_2$; n = 2, …, 21 | 61 |
| | 150 - 1200 | electrical self-heating | suspended and supported FLG; polymeric residues on the surface. | 68 |
| | 302-596 | modified T-bridge | suspended; n=2 – 8. | 69 |

| FLG nano-ribbons | 1100 | electrical self-heating | supported; exfoliat-ed; n<5 | 70 |
|---|---|---|---|---|
| | 80 - 150 | electrical self-heating | supported | 71 |

An experimental study [61] found $K \approx 160$ W/mK for encased single-layer graphene (SLG) at $T$=310 K. It increases to ~1000 W/mK for graphite films with the thickness of 8 nm. It was also found that the suppression of thermal conductivity in encased graphene, as compared to bulk graphite, was stronger at low temperatures where $K$ was proportional to $T^{\beta}$ with $1.5 < \beta < 2$ [61]. Thermal conduction in encased FLG was limited by the rough boundary scattering and disorder penetration through graphene.

Recently the measurements of $K$ in twisted bilayer graphene (T-BLG) were performed using an optothermal Raman technique [62]. The obtained values of $K = 1400 - 700$ W/mK in a temperature range $320 - 750$ K are almost by a factor of 2 smaller than in SLG and by a factor of 1.4 smaller than in Bernal-stacked bilayer graphene (BLG). The twisting affects phonon energy spectra, changes selection rules for phonon transitions and opens up new paths for phonon relaxation [62-64] (see figure 3). The experimental data on thermal conductivity in graphene and FLG is presented in Table 1.

### 3. Theories of phonon thermal conductivity in graphene, few-layer graphene and graphene nanoribbons

The first experimental investigations of the thermal properties in graphene materials [19-20, 22, 47-48, 55] stimulated numerous theoretical and computational works in the field. Here, we briefly review the state-of-the-art in theory of thermal transport in graphene and GNRs. Many different theoretical models have been proposed for the prediction of the phonon and thermal properties in graphite, graphene and GNRs during the last few years. The phonon energy spectra have been theoretically investigated using Perdew-Burke-Ernzerhof generalized gradient approximation (GGA) [72-74], valence-force-field (VFF) and Born-von Karman models of lattice vibrations [23-24, 26, 63, 75-76], continuum approach [77-79], first-order local density approximation [73,80,81], fifth- and fourth-nearest neighbor force constant approaches [74,82] or utilized the Tersoff, Brenner or Lennard-Jones potentials [59-60,83]. The thermal conductivity investigations have been performed within molecular dynamics simulations [57, 84-99], density functional theory [100, 101], Green's function method [102,103] and Boltzmann-transport-equation (BTE) approach [21-26, 35, 45, 46, 59, 60, 83, 104-107]. It has been shown that phonon energies strongly depend on the interatomic force constants (IFCs) – fitting parameters of interatomic interactions, used in the majority of the models. Therefore a proper choice of interatomic force constants is crucial for the accurate description of phonon energy spectra and thermal conductivity in graphene, twisted graphene and graphene nanoribbons [18, 27-28, 63].

Although various models predicted different values of thermal conductivity, they demonstrated consistent results on the strong dependence of graphene lattice thermal conductivity on extrinsic parameters of flakes: edge quality, FLG thickness, lateral size and shape, lattice strain, isotope, impurity and grain concentration. The molecular dynamic (MD) simulations give usually smaller values of thermal conductivity in comparison with BTE model and experimental data due to exclusion of long wavelength phonons from the model by a finite size of the simulation domain [27]. The effect of the edge roughness on the thermal conductivity in graphene and GNRs has been investigated in Refs. [21, 23-26, 45, 46, 79, 84, 104, 107-109]. The rough edge can suppress the thermal conductivity by an order of magnitude as compared to that in graphene or GNRs with perfect edges due to the boundary scattering of phonons. Impurities, single vacancies, double vacancies and Stone-Wales defects decrease the thermal conductivity of graphene and GNRs by more than 50% - 80% in dependence of the defect concentration [21, 23, 24, 26, 90-94].

A study of thermal conductivity of graphene and GNRs under strain was performed in Refs. [88, 100-103, 110]. An enhancement of the thermal conductivity of up to 36% for the strained 5-nm armchair or zigzag GNRs was found in the ballistic transport regime [103]. In the diffusive transport regime, the applied strain enhanced the Umklapp scattering and thermal conductivity diminishes by ~ 1.4 orders of magnitude at RT in comparison with the unstrained graphene [101]. The authors of Ref. [88] have found that when the strain is applied in both directions—parallel and perpendicular to the heat transfer path—the graphene sheets undergo complex reconstructions. As a result, some of the strained graphene structures can have higher thermal conductivity than that of SLG without strain [88]. The discrepancy between theoretical findings and experiments requires additional investigations of thermal transport in strained graphene and GNRs. The isotope composition is another key parameter for thermal conductivity engineering in these materials [18, 27-28, 111- 115]. Naturally occurring carbon materials are made up of two stable isotopes of 12C (∼99%) and 13C (∼1%). The change in the isotope composition significantly influences the crystal lattice properties. Increasing the "isotope doping" leads to a suppression of the thermal conductivity in graphene and GNRs of up to two orders of magnitude at RT due to the enhanced phonon-point defect (mass-difference) scattering [27, 106, 111-115].

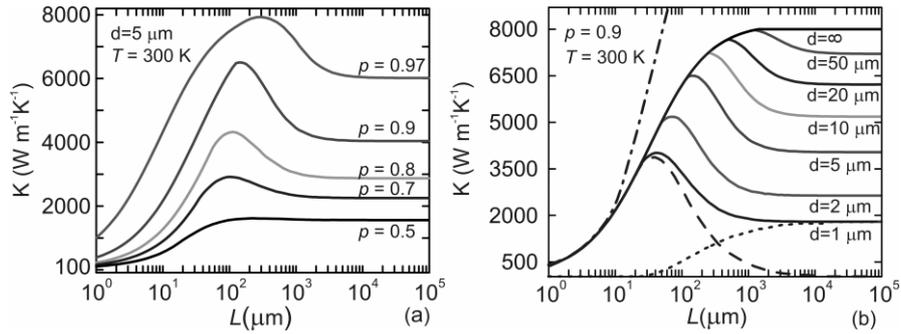

**Fig. 4.** (a) Dependence of the thermal conductivity of the rectangular graphene ribbon on the ribbon length $L$ shown for different specular parameters $p$. The ribbon width is fixed at $d=$ 5 μm. (b) Dependence of the thermal conductivity of the rectangular graphene ribbon on the ribbon length $L$ shown for different ribbon width $d$. The specular parameter is fixed at $p=0.9$. Note in both panels an unusual non-monotonic length dependence of the thermal conductivity, which results from the exceptionally long phonon mean free path of the low-energy phonons and their angle-dependent scattering from the ribbon edge. Figure is after Ref. [26] reproduced with permission from the American Chemical Society.

Graphene and GNRs also demonstrated an intriguing dependence of the thermal conductivity on their geometrical parameters: lateral sizes and shapes [23, 24, 26, 79, 95-98, 107]. Using BTE approach, Nika *et al.* [26] have demonstrated that RT thermal conductivity of a rectangular graphene flake with 5 μm width increases with length $L$ up to $L \sim 40 - 200$ μm and converges for $L > 50 - 1000$ μm in dependence on the phonon boundary scattering parameter $p$ (see Figure 4 (a)). The dependence of the thermal conductivity on $L$ is non-monotonic, which is explained by the interplay between contribution to the thermal conductivity from two groups of phonons: participating and non-participating in the edge scattering [26]. The exceptionally large mean free path (MFP) of the acoustic phonons in graphene is essential for this effect. The increase in the flake width or phonon edge scattering (see Figure 4 (a-b)) attenuates the non-monotonic behavior. It disappears in circular flakes or flakes with very rough edges (with specular parameter $p<0.5$).

A number of studies [95-97] employed the MD simulations to investigate the length dependence of the thermal conductivity in graphene and GNRs. The converged thermal conductivity in graphene was found for $L>16$ μm in Ref. [95]. In Refs. [96,97] the thermal conductivity increases monotonically with an increase of the length up to 2.8 μm in graphene [97] and 800 nm in GNRs [96]. The obvious length dependence in graphene and GNRs can be attributed to the extremely large phonon mean free path $\Lambda \sim 775$ nm [20], which provides noticeable length dependence even for flakes with micrometer lengths.

Keblinsky and co-workers [84] found from the MD study that the thermal conductivity of graphene is $K \approx 8000 - 10000$ W/mK at RT for the square graphene sheet. The $K$ value was size independent for $L>5$ nm [84]. For the ribbons with fixed $L=10$ nm and width $W$ varying from 1 to 10 nm, $K$ increased from ~1000 W/mK to 7000 W/mK. The thermal conductivity in GNR with rough edges can be suppressed by orders of magnitude as compared to that in GNR with perfect edges [84, 108]. The isotopic superlattice modulation of GNR or defects of crystal lattices also significantly decreases the thermal conductivity [111, 116]. The uniaxial stretching applied in the longitudinal direction enhances the low-temperature thermal conductance for the 5 nm arm-chair or zigzag GNR up to 36 % due to the stretching-induced convergence of phonon spectra to the low-frequency region [103].

Aksamija and Knezevic [104] calculated the dependence of the thermal conductivity of GNR with the width 5 nm and RMS edge roughness $\Delta = 1$ nm on temperature. The thermal conductivity was calculated taking into account the three-phonon Umklapp, mass-defect and rough edge scatterings [104]. The authors obtained RT thermal conductivity $K \sim 5500$ W/mK for the graphene nanoribbon. The study of the nonlinear thermal transport in rectangular and triangular GNRs under the large temperature biases was reported in Ref. [117]. The authors found that in short (~6 nm) rectangular GNRs, the negative differential thermal conductance exists in a certain range of the applied temperature difference. As the length of the rectangular GNR increases the effect weakens. A computational study reported in Ref. [118] predicted that the combined effects of the edge roughness and local defects play a dominant role in determining the thermal transport properties of zigzag GNRs. The experimental data on thermal transport in GNRs is very limited. In Ref. [70] the authors used an electrical self-heating methods and extracted the thermal conductivity of sub 20-nm GNRs to be more than 1000 W/mK at $700 - 800$ K. A similar experimental method but with more accurate account of GNRs thermal coupling to the substrate has been used in Ref. [71]. Pop and co-workers [71] found substantially lower values of thermal conductivity of ~ $80 - 150$ W/mK at RT.

Ong and Pop [87] examined thermal transport in graphene supported on $SiO_2$ using MD simulations. The approach employed by the authors utilized the reactive empirical bond order (REBO) potential to model the atomic interaction between the C atoms, Munetoh potential to model the atomic interactions between the Si and O atoms and Lennard-Jones potential to model the van der Waals type C-Si and C-O couplings. Authors suggested that thermal conductivity in supported graphene is by an order of magnitude smaller than that in suspended graphene due to damping of the out-of-plane ZA phonons.

**Table 2.** Thermal conductivity of graphene and few-layer graphene: theoretical data.

| Sample | K (W/mK) | Method | Description | Ref |
|--------|----------|--------|-------------|-----|
| **SLG** | 1000 - 8000 | BTE, $\gamma_{LA}$, $\gamma_{TA}$ | strong size dependence | 24 |
| | 2000-8000 | BTE, $\gamma_s(q)$ | strong edge, width and grunaisen parameter dependence | 23 |
| | ~2430 | BTE, | $K$(graphene) $\geq K$ (carbon | 120 |
| | | $3^{rd}$-order interatomic force constants (IFCs) | nanotube) | |
| | 1500 - 3500 | BTE, $3^{rd}$-order IFCs | strong size dependence | 59 |
| | 100 - 8000 | BTE | Strong length, size, shape and edge dependence. | 26 |
| | 2000 - 4000 | continuum approach + BTE | strong isotope, point-defects and strain influence. | 79, 121 |
| | ~ 4000 | ballistic | strong width dependence | 122 |
| | ~ 2900 | MD simulation | strong dependence on the vacancy concentration | 85 |
| | ~ 20000 | VFF + MD simulation | Ballistic regime; flake length ~ 5 μm; strong width and length dependence | 123 |
| | 100-550 | MD simulation | flake length $L$<200 nm; strong length and defect dependence | 92 |
| | ~ 3000 | MD simulation | sheet length ~ 15 μm; strong size dependence | 95 |
| | 2360 | MD simulation | $L$~5 μm; strong length dependence | 97 |
| | 4000-6000 | MD simulation | strong strain dependence | 101 |
| | ~ 3600 | Boltzmann-Peierls equation + density functional perturbation theory | $L=10$ μm; insensitivity to small isotropic strain | 124 |
| | ~ 1250 | MD simulation | $L$=100 μm; strong length dependence for $L$<100 μm | 125 |
| | 1800 | MD simulation | 6 nm × 6 nm sheet; isolated | 99 |
| | 1000-1300 | MD simulation | 6 nm × 6 nm sheet; Cu − supported; strong dependence on the interaction strength between graphene and substrate | |
| **FLG** | 1000 - 4000 | BTE, $\gamma_s(q)$ | n = 8 − 1, strong size dependence | 22 |

| | 1000 - 3500 | BTE, $3^{rd}$-order IFCs | n = 5 − 1, strong size dependence | 59 |
|---|---|---|---|---|
| | 2000-3300 | BTE, $3^{rd}$-order IFCs | n = 4 − 1 | 60 |
| | 580 - 880 | MD simulation | n = 5 − 1, strong dependence on the Van-der Vaals bond strength | 86 |

**Table 3.** Thermal conductivity of GNRs: theoretical data.

| K (W/mK) | Method | Description | Ref |
|---|---|---|---|
| 1000 - 7000 | Theory: molecular dynamics, Tersoff | strong ribbon width and edge dependence | 84 |
| ~ 5500 | BTE | GNR with width of 5 μm; strong dependence on the edge roughness | 104 |
| ~2000 | MD simulation | $T$=400 K; 1.5 nm × 5.7 nm zigzag GNR; strong edge chirality influence | 109 |
| 30-80 | AIREBO potential + MD simulation | 10 - zigzag and 19 -armchair nanoribbons; strong defect dependence | 91,93 |
| 3200-5200 | MD simulation | strong GNRs width ($W$) and length dependence; 9 nm ≤ $L$ ≤27 nm and 4 nm ≤ $W$ ≤18 nm | 94 |
| 400 - 600 | MD simulation | $K \sim L^{0.24}$; 100 nm ≤ $L$ ≤ 650 nm | 96 |
| 100 - 1000 | BTE | GNRs supported on SiO₂; strong edge and width dependence | 107 |
| 500 - 300 | MD simulation | few-layer GNRs; 10-ZGNR, $n = 1,...,5$ | 98 |

The strong dependence of the thermal conductivity of graphene on the defect concentration was established in the computational studies reported in Refs. [85, 89]. Both studies used MD simulations. According to Hao et al. [89] 2 % of the vacancies or other defects can reduce the thermal conductivity of graphene by as much as a factor of five to ten. Zhang et al. [85] determined from their MD simulations that the thermal conductivity of pristine graphene should be ~2903 W/mK at RT. According to their calculations the thermal conductivity of graphene can be reduced by a factor of 1000 at the vacancy defect concentration of ~9 %. The numeric results of Refs.

[85, 89] suggest another possible explanation of the experimental data scatter, which is different defect density in the examined graphene samples. For example, if the measurements of the thermal conductivity of graphene by the thermal bridge technique give smaller values than those by the Raman optothermal technique, one should take into account that the thermal bridge technique requires substantial number of fabrication steps, which result in residual defects.

The available theoretical values of phonon thermal conductivity in SLG, few-layer graphene and GNRs are presented in Tables 2 and 3 at RT (if not indicated otherwise). Readers interested in a more detailed description of theoretical models for the heat conduction in graphene materials are referred to review papers [27, 28, 119].

### 4.1. Specifics of two-dimensional phonon transport

We now address in more detail some specifics of the acoustic phonon transport in 2D systems. Investigation of the heat conduction in graphene [19-20] and CNTs [126] raised the issue of ambiguity in the definition of the intrinsic thermal conductivity for 2D and 1D crystal lattices. It was theoretically shown that the intrinsic thermal conductivity limited by the crystal anharmonicity has a finite value in 3D bulk crystals [12, 56]. However, many theoretical models predict that the intrinsic thermal conductivity reveals a logarithmic divergence in strictly 2D systems, $K \sim ln(N)$, and the power-law divergence in 1D systems, $K \sim N^{\alpha}$, with the number of atoms $N$ ($0 < \alpha < 1$) [12, 16, 56, 126-130]. The logarithmic divergence can be removed by introduction of the *extrinsic* scattering mechanisms such as scattering from defects or coupling to the substrate [56]. Alternatively, one can define the *intrinsic* thermal conductivity of a 2D crystal for a given size of the crystal.

Graphene is not an ideal 2D crystal, considered in most of the theoretical works, since graphene atoms vibrate in three directions. Nevertheless, the intrinsic graphene thermal conductivity strongly depends on the graphene sheet size due to weak scattering of the low-energy phonons by other phonons in the system. Therefore, the phonon boundary scattering is an important mechanism for phonon relaxation in graphene. Different studies [26,131-132] also suggest-ed that an accurate accounting of the higher-order anharmonic processes, i.e. above three-phonon Umklapp scattering, and inclusion of the normal phonon processes into consideration allow one to limit the low-energy phonon MFP. The normal phonon processes do not contribute directly to thermal resistance but affect the phonon mode distribution [59, 120]. However, even these studies found that the graphene sample has to be very large (>10 μm) to obtain the size-independent thermal conductivity.

In BTE approach within relaxation time approximation the thermal conductivity in cvuasi-2D system are given by [23, 26]:

$$
K = \frac{1}{4\pi k_B T^2 h} \times
$$
$$
\times \sum_s \int_0^{q_{max}} \{ [\hbar\omega_s(q) \frac{d\omega_s(q)}{dq}]^2 \tau_{tot}(s,q) \frac{\exp[\hbar\omega_s(q)/kT]}{[\exp[\hbar\omega_s(q)/kT]-1]^2} q\}dq. \tag{7}
$$

Here $\hbar\omega_s(q)$ is the phonon energy, $h = 0.335$ nm is the graphene layer thickness, $\tau_{tot}$ is the total phonon relaxation time, $q$ is the phonon wavenumber, $T$ is the temperature and $k_B$ is the Boltzmann constant.

The specific phonon transport in the quasi - 2D system such as graphene can be illustrated with a simple expression for Umklapp – limited thermal conductivity derived by us in Ref. [24]:

$$
K_U = \frac{M}{4\pi Th} \sum_{s=TA,LA} \frac{\omega_{s,max} \overline{U}_s^2}{\gamma_s^2} F(\omega_{s,min}, \omega_{s,max}), \tag{8}
$$

where

$$
F(\omega_{s,min}, \omega_{s,max}) = \int_{\hbar\omega_{s,min}/k_B T}^{\hbar\omega_{s,max}/k_B T} \xi \frac{exp(\xi)}{[exp(\xi)-1]^2} d\xi =
$$
$$
[ln\{exp(\xi)-1\} + \frac{\xi}{1-exp(\xi)} - \xi]_{\hbar\omega_{s,min}/k_B T}^{\hbar\omega_{s,max}/k_B T}. \tag{9}
$$

In the above equation, $\xi = \hbar\omega / k_B T$, and the upper cut-off frequencies $\omega_{s,max}$ are defined from the actual phonon dispersion in graphene [23]: $\omega_{LA,max} = 2\pi f_{LA,max} (\Gamma K) = 241$ rad/ps, $\omega_{TA,max} = 2\pi f_{TA,max} (\Gamma K) = 180$ rad/ps. The integrand in equation (9) can be further simplified near RT when $\hbar\omega_{s,max} > k_B T$, and it can be expressed as

$$F(\omega_{s,min}) \approx -ln\{|\,exp(\hbar\omega_{s,min} / k_B T) - 1\,|\} +$$
$$+ \frac{\hbar\omega_{s,min}}{k_B T} \frac{exp(\hbar\omega_{s,min} / k_B T)}{exp(\hbar\omega_{s,min} / k_B T) - 1} \qquad (10)$$

In Eqs [7-10] the contribution of ZA phonons to thermal transport has been neglected [24, 35, 133] because of their low group velocity and large Gruneisen parameter $\gamma_{ZA}$ [24, 73].

There is a clear difference between the heat transport in basal planes of bulk graphite and in single layer graphene [35, 133]. In the former, the heat transport is approximately two-dimensional only up to some lower-bound cut-off frequency $\omega_{min}$. Below $\omega_{min}$ there appears to be strong coupling with the cross-plane phonon modes and heat starts to propagate in all directions, which reduces the contributions of these low-energy modes to heat transport along basal planes to negligible values. In bulk graphite, there is a physically reasonable reference point for the on-set of the cross-plane coupling, which is the ZO' phonon branch near ~4 THz observed in the spectrum of bulk graphite [35, 134]. The presence of the ZO' branch and corresponding $\omega_{min} = \omega_{ZO'} (q = 0)$ allows one to avoid the logarithmic divergence in the Umklapp-limited thermal conductivity integral [see equations (7–10)] and calculate it without considering other scattering mechanisms.

The physics of heat conduction is principally different in graphene where the phonon transport is 2D all the way to zero phonon frequency. Therefore the lower-bound cut-off frequencies $\omega_{s,min}$ for each $s$ are determined from the condition that the phonon MFP cannot exceed the physical size $L$ of the flake, i.e.

$$\omega_{s,\min} = \frac{\bar{\upsilon}_s}{\gamma_s} \sqrt{\frac{M\bar{\upsilon}_s}{k_B T} \frac{\omega_{s,\max}}{L}} \ . \tag{11}$$

We would like to emphasize here that using size-independent graphite $\omega_{\min}$ for SLG or FLG (as has been proposed in Ref. [135]) is without scientific merit and leads to an erroneous calculation of thermal conductivity, as described in detail in Ref. [25]. Equations (8-11) constitute a simple analytical model for the calculation of the thermal conductivity of the graphene layer, which retains such important features of graphene phonon spectra as different $\bar{\upsilon}_s$ and $\gamma_s$ for *LA* and *TA* branches. The model also reflects the two-dimensional nature of heat transport in graphene all the way down to zero phonon frequency.

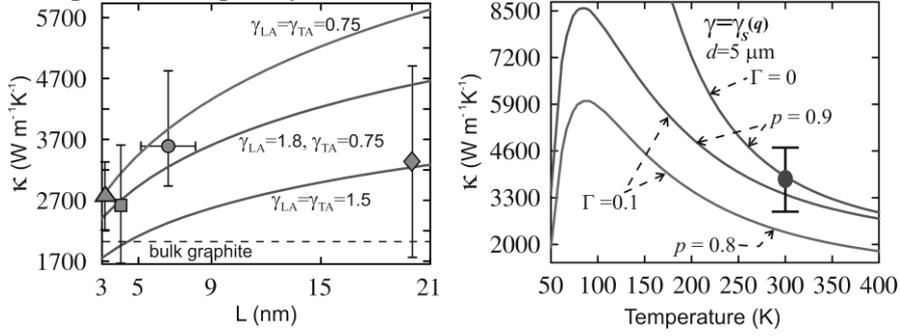

**Fig. 5.** (a) Calculated room temperature thermal conductivity of graphene as a function of the lateral size for several values of the Gruneisen parameter. (b) Calculated thermal conductivity of suspended graphene as a function of the temperature. Note a strong dependence on the size of the graphene flakes. Experimental data points from Refs. [19-20] (circle), [47] (square), [48] (rhomb) and [62] (triangle) are shown for comparison. Figure 5(b) is after Ref. [23] reproduced with permission from the American Physical Society.

In figure 5(a), we present the dependence of thermal conductivity of graphene on the dimension of the flake *L*. The data is presented for the averaged values of the Gruneisen parameters $\gamma_{LA}$=1.8 and $\gamma_{TA}$=0.75 obtained from *ab initio* calculations, as well as for several other close sets of $\gamma_{LA,TA}$ to illustrate the sensitivity of the result to the Gruneisen parameters. For small graphene flakes, the *K* dependence on *L* is rather strong. It weakens for flakes with $L\geq10$ μm. The calculated values are in good agreement with experimental data for suspended exfoliated [19-20] and CVD graphene [47-48]. The hori-zontal dashed line indicates the experimental thermal conductivity for bulk graphite, which is exceeded by graphene's thermal conductivity at smaller $L$. Thermal conductivity, presented in figure 5, is an *intrinsic* quantity limited by the three-phonon Umklapp scattering only. But it is determined for a specific graphene flake size since $L$ defines the lower-bound (long-wavelength) cut-off frequency in Umklapp scattering through equation (11). In experiments, thermal conductivity is also limited by defect scattering. When the size of the flake becomes very large with many polycrystalline grains, the scattering on their boundaries will also lead to phonon relaxation. The latter can be included in this model through adjustment of $L$. The extrinsic phonon scattering mechanisms or high-order phonon-phonon scatterings prevent indefinite growth of thermal conductivity of graphene with $L$ [26].

The simple model described above is based on the Klemens-like expressions for the relaxation time (see Refs. [35-36]). Therefore it does not take into account all peculiarities of the 2D three-phonon Umklapp processes in SLG or FLG, which are important for the accurate description of thermal transport. There are two types of the three-phonon Umklapp scattering processes [23, 36]. The first type is the scattering when a phonon with the wave vector $\vec{q}(\omega)$ absorbs another phonon from the heat flux with the wave vector $\vec{q}'(\omega')$, i.e. the phonon leaves the state $\vec{q}$. For this type of scattering processes the momentum and energy conservation laws are written as:

$$\begin{aligned} \vec{q}(\omega) + \vec{q}'(\omega') &= \vec{b}_i + \vec{q}''(\omega''), \; i = 1, 2, 3 \\ \omega + \omega' &= \omega'' \end{aligned}. \tag{12}$$

The processes of the second type are those when the phonons $\vec{q}(\omega)$ of the heat flux decay into two phonons with the wave vectors $\vec{q}'(\omega')$ and $\vec{q}''(\omega'')$, i.e. leaves the state $\vec{q}(\omega)$, or, alternatively, two phonons $\vec{q}'(\omega')$ and $\vec{q}''(\omega'')$ merge together forming a phonon with the wave vector $\vec{q}(\omega)$, which correspond to the phonon coming to the state $\vec{q}(\omega)$. The conservation laws for this type are given by:

$$\vec{q}(\omega) + \vec{b}_i = \vec{q}'(\omega') + \vec{q}''(\omega''), \quad i = 4, 5, 6$$
$$\omega = \omega' + \omega'', \tag{13}$$

In equations (12-13) $\vec{b}_i = \overrightarrow{\Gamma \Gamma_i}$, $i = 1, 2, ..., 6$ is one of the vectors of the reciprocal lattice. Calculations of the thermal conductivity in graphene taking into account all possible three-phonon Umklapp processes allowed by the equations (12-13) and actual phonon dispersions were carried out in Ref. [23]. For each phonon mode ($q_i$, $s$), were found all pairs of the phonon modes ($\vec{q}'$, $s'$) and ($\vec{q}''$, $s''$) such that the conditions of equations (12-13) are met. As a result, in ($\vec{q}'$)-space were constructed the *phase diagrams* for all allowed three-phonon transitions [23]. Using the long-wave approximation for a matrix element of the three-phonon interaction one can obtain for the Umklapp scattering rates:

$$\frac{1}{\tau_U^{(I),(II)}(s,\vec{q})} = \frac{\hbar \gamma_s^2(\vec{q})}{3\pi \rho \upsilon_s^2(\vec{q})} \times$$
$$\times \sum_{s's'';\vec{b}_i} \iint \omega_s(\vec{q}) \omega_{s'}'(\vec{q}') \omega_{s''}''(\vec{q}'') \times \{N_0[\omega_{s'}'(\vec{q}')] \mp \tag{14}$$
$$\mp N_0[\omega_{s''}''(\vec{q}'')] + \frac{1}{2} \mp \frac{1}{2}\} \times \times \delta[\omega_s(\vec{q}) \pm \omega_{s'}'(\vec{q}') - \omega_{s''}''(\vec{q}'')] dq_l' dq_\perp'.$$

Here $q_l'$ and $q_\perp'$ are the components of the vector $\vec{q}'$ parallel or perpendicular to the lines defined by equations (12-13), correspondingly, $\gamma_s(\vec{q})$ is the mode-dependent Gruneisen parameter, which is determined for each phonon wave vector and polarization branch and $\rho$ is the surface mass density. In equation (14) the upper signs correspond to the processes of the first type while the lower signs correspond to those of the second type. The integrals for $q_l, q_\perp$ are taken along and perpendicular to the curve segments, correspondingly, where the conditions of equations (12-13) are met.

The main mechanisms of phonon scattering in graphene are phonon-phonon Umklapp (U) scattering, rough edge scattering (boundary (B)) and point-defect (PD) scattering:

$$\frac{1}{\tau_{tot}(s,q)} = \frac{1}{\tau_U(s,q)} + \frac{1}{\tau_B(s,q)} + \frac{1}{\tau_{PD}(s,q)}, \qquad (15)$$

where $1/\tau_U = 1/\tau_U^I + 1/\tau_U^{II}$, $1/\tau_B(s,q) = (\upsilon_s/L)((1-p)/(1+p))$ and $1/\tau_{PD}(s,q) = S_0\Gamma q_s\omega_s^2/(4\upsilon_s)$. Here $\upsilon_s = d\omega_s/dq$ is the phonon group velocity, $p$ is the specularity parameter of rough edge scattering, $S$ is the surface per atom and $\Gamma$ is the measure of the strength of the point defect scattering.

The sensitivity of the thermal conductivity, calculated using equations (7, 12-15), to the value of $p$ and $\Gamma$ is illustrated in figure 5(b). The data is presented for different sizes (widths) of the graphene flakes.

## 4. Conclusions

We reviewed theoretical and experimental results pertinent to 2D phonon transport in graphene. Phonons are the dominant heat carriers in the ungated graphene samples near room temperature. The unique nature of 2D phonons translates to unusual heat conduction in graphene and related materials. Recent computational studies suggest that the thermal conductivity of graphene depends strongly on the concentration of defects and strain distribution. Investigation of the physics of 2D phonons in graphene can shed light on the thermal energy transfer in low-dimensional systems. The results presented in this review are important for the proposed practical applications of graphene in heat removal and thermal management of advanced electronics.

**Acknowledgments**

The work at UC Riverside was supported, in part, by the National Science Foundation (NSF) project CMMI-1404967 Collaborative Research Genetic Algorithm Driven Hybrid Computational Experimental Engineering of Defects in Designer Materials; NSF project ECCS-1307671 Two-Dimensional Performance with Three-Dimensional Capacity: Engineering the Thermal Properties of Graphene, and by the STARnet Center for Function Accelerated nanoMaterial Engineering (FAME) – Semiconductor Research Corporation (SRC) program sponsored by The Microelectronics Advanced Research Corporation (MARCO) and the Defense Advanced Research Project Agency (DARPA). The work at Moldova State University was supported, in part, by the Moldova State Project 15.817.02.29F and ASM-STCU project #5937.